\documentclass{article}

\usepackage{amsmath}
\usepackage{PRIMEarxiv}
\usepackage[numbers]{natbib}
\usepackage[utf8]{inputenc} % allow utf-8 input
\usepackage[T1]{fontenc}    % use 8-bit T1 fonts
\usepackage{hyperref}       % hyperlinks
\usepackage{url}            % simple URL typesetting
\usepackage{booktabs}       % professional-quality tables
\usepackage{amsfonts}       % blackboard math symbols
\usepackage{nicefrac}       % compact symbols for 1/2, etc.
\usepackage{microtype}      % microtypography
\usepackage{lipsum}
\usepackage{fancyhdr}       % header
\usepackage{graphicx}       % graphics

\title{Accelerating Posterior sampling for Scalable Gaussian Process model
}

\author{
  Zhihao, Zhou \\
  Keck School of Medicine \\
  University of Southern California \\
  Los Angeles\\
  \texttt{zhouhih@usc.edu} \\
}

\begin{document}
\maketitle

\begin{abstract}
This thesis conducts a thorough simulation study to assess the effectiveness of various acceleration techniques designed to enhance the conjugate gradient algorithm, which is used for solving large linear systems to accelerate Bayesian computation in spatial analysis. The focus is on the application of symbolic decomposition and preconditioners, which are essential for the computational efficiency of conjugate gradient. The findings reveal notable differences in the effectiveness of these acceleration methods. Specific preconditioners, such as the Diagonal Preconditioner, consistently delivered improvements in computational speed. However, in settings involving high-dimensional matrices, traditional solvers were less effective, underscoring the importance of specialized acceleration techniques like the diagonal preconditioner and cgsparse. These methods demonstrated robust performance across a variety of scenarios. The results of this study not only enhance our understanding of the algorithmic dynamics within spatial statistics but also offer valuable guidance for practitioners in choosing the most appropriate computational techniques for their specific needs.
\end{abstract}

% keywords can be removed

\section{GAUSSIAN PROCESS BASED SPATIAL REGRESSION MODEL}
This chapter presents background and essential element of the Gaussian process-based spatial model for point-referenced data. 
\subsection{Background}
The increasing adoption and evolving functionalities of Geographic Information Systems (GIS) have led to a surge in research focused on the modeling and analysis of spatial data in various fields, including but not limited to environmental science, economics, and biometry. This broad interest underscores the increasing relevance of GIS in navigating the complexities of spatial data interpretation and management. Spatial modeling often employs the hierarchical modeling framework, formulated as 

$$[data | process]*[process| parameters]*[parameters]$$ 

In the context of point-referenced data, where spatial locations are denoted by map coordinates, the process is represented by a spatial random field across the domain, with observations considered as finite realizations of this field. The Gaussian process (GP) stands out as a notably versatile process specification, enabling flexible and comprehensive modeling. Its application is widespread, benefiting from its adaptability to both multivariate scenarios and spatial-temporal geostatistical models. Despite its advantages, including theoretical tractability, the implementation of GPs (a series of Gaussian processes) for large datasets is computationally demanding. This is primarily due to the substantial spatial covariance matrix required in the multivariate normal density for GP realizations. For spatial data points that are irregularly distributed (which is a common scenario in geostatistics), these matrices tend to be dense and lack a structure that could ease computational efforts. Consequently, even for datasets with a relatively large number of points (approximately 50,000 or more), the computational requirements can exceed the capacities of contemporary computers, hindering practical applications of GP models.

Fitting Gaussian Process (GP) models involves significant computational expenses, limiting their use for extensive datasets. The primary issue originates from the extensive spatial covariance matrix needed to model spatial correlations among data points. This is especially challenging for spatially irregular locations, typical in geostatistics, where these matrices are generally dense without any structure to simplify computations. The requisite Cholesky decomposition for likelihood evaluations, whose storage and computational demands grow quadratically and cubically with the location count, quickly becomes unmanageable with increasing data size. For datasets around 50,000 points, the computational needs surpass what modern computing can handle, obstructing inference from GP models.

Bypassing the computationally demanding tasks like matrix multiplication and factorization in the optimization of GP model hyperparameters is feasible by solving linear systems with large covariance matrices. The Conjugate Gradient (CG) method, utilized in significant research for parameter estimation, evidences this approach. For instance, Stein\cite{Stein2012}. investigated preconditioning methods for the covariance matrix in score equations, implementing iterative algorithms for estimating hyperparameters. Sun and Stein\cite{Sun2016} applied preconditioned CG to achieve maximum likelihood hyperparameter estimation. Furthermore, CG's application in simulating exact GP models is highlighted by Wang\cite{Wang2019}'s use of multi-GPU parallelization. CG's role in Bayesian spatial analysis is chiefly in enhancing posterior sampling procedures, as demonstrated by Stroud\cite{Stroud2017}'s two-block Gibbs sampler and Zhang and Banerjee\cite{Zhang2021}'s scalable spatial factor models, underscoring CG's wide applicability in spatial data analysis.
\subsection{GP for point-referenced Spatial Data}
Spatial modeling for point-referenced data heavily relies on GP. 
 GP are flexible non-parametric models, with a capacity that grows with the available data. GPs have demonstrated remarkable effectiveness in various machine learning applications, including black-box optimization\cite{Snoek2012}, reinforcement learning\cite{Deisenroth2011,Deisenroth2015}, and time series forecasting \cite{Roberts2013}. These models bring numerous benefits, such as well-founded representations of uncertainty, minimal need for expert-driven model priors, and scalable adaptability to datasets of any size. GPs excel in scenarios with limited observations and show significant potential in harnessing insights from large datasets, particularly when integrated with expressive kernels or hierarchical frameworks.

\subsection{GP based Spatial Regression model}
A broader perspective considers a spatial regression model applicable to any location $s$
\begin{equation}
y(s)=m_\theta+\omega(s)+\epsilon(s),\epsilon(s) \stackrel{\text{iid}}{\sim} N(0,\tau^2)
\end{equation}
where, usually, $m_\theta(s)=x^T(s)\beta$ and $\omega(s)$ is a latent spatial process capturing spatial dependence. 
We denote the outcome of interest to be $Y(s)$, where $D\subset R^d$ is a fixed subset of $\mathbb{R}^d$. Assuming that there will be $p$ predictors for each location, we denote $Y(s)$ as the corresponding $p$ predictors observed at the location $s$. Then, the customarily linear regression model can be formatted as 
\begin{equation}\label{eq: spatial_model}
y(s)=x^T(s) \beta + \omega(s)+\epsilon(s)
\end{equation}
where the $\omega(s)$ stands for the spatial random effect across the whole space. And $\epsilon(s)$ is the white noise that is independent and identically distributed across different locations and follow the distribution of $N(0,\tau^2)$ with $\tau^2$ being the variance of the white noise. The $\omega(s)$ here is used to model the underlined spatial pattern of our outcome and is customarily modeled by the Gaussian process. 
Here, $\omega(s)$ is hierarchically modeled by a Gaussian process with a prior for $\beta$, $\tau^2$ and $\phi$.

\subsection{Challenges for posterior of high-dimension latent process}
Fitting GP models presents significant computational challenges, especially when managing large datasets. This difficulty mainly arises from the extensive spatial covariance matrix needed to encapsulate the spatial correlations among observations. In geostatistical applications, where spatial locations often are irregularly distributed, this covariance matrix is generally dense, lacking any inherent structure that could otherwise ease computational efforts.

The computational and storage demands of performing Cholesky decomposition, essential for likelihood evaluation within GP models, escalate sharply with the number of spatial locations. Specifically, if data are observed in $n$ distinct locations, the storage space needed for the spatial covariance matrix scales to $\mathcal{O}(n^2)$, while the computational expense for evaluating the likelihood rises to $\mathcal{O}(n^3)$.

Given these scaling factors, even datasets with a moderately large number of points - around 50,000 or more — can overwhelm the capabilities of contemporary computing systems, rendering the direct application of GP models infeasible for extensive spatial data analysis. The computational burden becomes more onerous when generating full Bayesian inference through iterative algorithms such as Markov chain Monte Carlo (MCMC), especially when each iteration computation involves likelihood evaluation. Instead of resorting to MCMC sampling algorithms, we can implement a conjugate (conditional) Bayesian framework to accelerate the posterior sampling for spatial regression models. This framework can be combine with scalable univariate spatial process models, which will be introduced in the next Chapter. 

\section{Conjugate scalable Spatial regression model}
\subsection{Conjugate Bayesian framework}
Consider a spatial linear regression model given by
\begin{equation}
y_s = X\beta + w_s + \epsilon_s, 
\end{equation}
where $y_s$, $w_s$, and $\epsilon_s$ denote the realizations of the respective stochastic processes across the observation locations $S = \{s_1, s_2, \ldots, s_n\}$. Here, $X$ represents an $n \times p$ design matrix whose $i$-th row corresponds to a $1 \times p$ vector of explanatory variables, $x(s_i)^\top$, for each location $s_i$ in $S$. We will henceforward omit explicit mention of the dependency on $S$ for $y_s$, $w$, $\epsilon$, and their covariance structures to avoid ambiguity.
% y --> y_s
Assuming Gaussian distributions, $w \sim \mathcal{N}(0,\sigma^2 C)$ and $\epsilon \sim \mathcal{N}(0,\delta^2\sigma^2 I)$, with known parameters $C$ and $\delta^2 = \tau^2 / \sigma^2$, we define $\gamma^\top = [\beta^\top,w^\top]$ and $\mu_\gamma^\top = [\mu_\beta^\top,O^\top]$ and $V_\gamma = \left[ \begin{array}{cc}
V_\beta & O \\
O& C
\end{array} \right]$. $O$ represents zero matrices distinguishing independent components within the prior covariance structure. The Normal-Inverse-Gamma (NIG) distribution serves as a suitable conjugate prior,
\begin{equation}
p(\gamma,\sigma^2) = \mathcal{NIG}(\gamma,\sigma^2 | \mu_\gamma, \Sigma_\gamma, a, b) = \mathcal{N}(\gamma | \mu, \sigma^2 V) \cdot \mathcal{IG}(\sigma^2 | a, b).
\end{equation}
The conjugate nature of the NIG distribution enables straightforward sampling from the posterior distribution, a cornerstone for exact Bayesian inference. The matrix $\Sigma_\gamma$ represents the covariance structure of the combined parameter vector $\gamma$, integrating both the regression coefficients and spatial random effects within the NIG prior framework. $a$ and $b$ define the shape and scale of the Inverse-Gamma prior for $\sigma^2$. The proportional posterior distribution is then expressed as
\begin{equation}\label{eq: posterior}
p(\gamma, \sigma^2 | y_s) \propto \mathcal{NIG}(\gamma, \sigma^2 | \mu, V, a, b) \cdot \mathcal{N}(y_s | [X : I]\gamma, \delta^2\sigma^2 I).
\end{equation}

The combined posterior adopts a $\mathcal{NIG}(\mu^*, V^*, a^*, b^*)$ form, where
\begin{align*}
\mu^* &= V^{*-1}(V^{-1}\mu + X^{*\top}y_s), \\
V^* &= [V^{-1} + X^{*\top}X^*]^{-1}, \\
a^* &= a + \frac{n}{2}, \\
b^* &= b + \frac{1}{2}(\mu^\top V^{-1} \mu + y_s\top y_s - \mu^{*\top}V^{*-1}\mu^*).
\end{align*}

With $\beta$'s prior modeled as $\mathcal{N}(\mu_\beta, V_\beta)$ and in the scenario of noninformative priors, the precision matrix for the $\gamma$'s prior simplifies to a null matrix, indicating the absence of prior information influencing the posterior, allowing us to set $\mu_\gamma^\top = [O^\top, O^\top]$.

The marginal posterior for $\sigma^2$ adheres to a distribution $\mathcal{IG}(a^*, b^*)$, while the marginal posterior $\gamma$' is represented by a multivariate $t$ distribution with parameters $\mu^*$, $b^*V^*$ and degrees of freedom $2a^*$. Because of the conjugacy, exact Bayesian inference can proceed via direct sampling from the joint posterior density, eliminating the need for iterative sampling methods such as MCMC. This facilitates efficient posterior sampling, especially for $\sigma^2$, from which we can derive samples for $\tau^2$ by the known transformation $\delta^2$, significantly streamlining the inference process within the conjugate Bayesian framework.

\subsection{Scalable Gaussian Process}
Investigating
\begin{equation}
X^{*\top}X^* = \left[ \begin{array}{cc}
\frac{1}{\delta^2}X^\top X + L_\beta^{-\top}L_\beta^{-1} & \frac{1}{\delta^2}X^\top \\
\frac{1}{\delta^2}X & A^{**}
\end{array} \right],
\end{equation}
In the preceding analysis, we observe that the matrix $A^{**}$ can be succinctly represented as
\begin{equation}
    A^{**} = C^{-1} + \frac{1}{\delta^2} I_n\;,
\end{equation}
where $C^{-1}$ is the precision matrix of the latent process $w(s)$ across the spatial domain $S$. Addressing the computational challenge in Gaussian Process (GP) models centers around managing the $n \times n$ covariance matrix $C_{\theta}(S,S)$, particularly for a substantial number of observed locations $S = \{s_1, s_2, \ldots, s_n\}$. To mitigate computational intensity, a practical approach involves substituting $C_{\theta}(S,S)$ with a sparser approximate matrix $\tilde{C}_{\theta}(S,S)$. Among a range of strategies, Gaussian Markov Random Fields (GMRFs) stand out for their ability to create sparse approximations conducive to computational efficiency\cite{Rue2009}. Furthermore, adopting methodologies similar to graphical models or Bayesian networks, as explained by Lauritzen\cite{Lauritzen1996}, Bishop\cite{Bishop2006}, and Murphy\cite{Murphy2012}, enables the formation of composite likelihoods for inference. This framework was further refined by Datta\cite{Heaton2017} with the innovation of the Nearest Neighbor Gaussian Process (NNGP), specifically designed for extensive spatial datasets. NNGP, by using its finite-dimensional Gaussian distributions and sparse precision matrices, significantly enhances the scalability of parameter estimation and spatial prediction or ''kriging''.

Assuming that the latent process $w(s)$ is modeled through the NNGP with each location $s$ having $m$ neighbors, this model captures the essential spatial dependencies with a significantly reduced computational burden. Specifically, the precision matrix $C^{-1}$ of the process is represented as $(I_n - A_M)^\top D^{-1}(I_n - A_M)$. Here, $A_M$ is identified as a sparse and strictly lower triangular matrix, containing no more than $m$ nonzero entries per row, which dramatically limits its computational complexity. $D$, a diagonal matrix, further simplifies the structure, facilitating efficient large-scale spatial analysis and enabling more manageable computations for large datasets, which are essential in efficiently handling the latent process.

Direct sampling from the posterior distribution in (\ref{eq: posterior}) is achieved by first sampling $\sigma^2 \sim \mbox{IG}(a^\ast, b^\ast)$ and then sampling one draw of $\gamma \sim \mbox{N}(\mu^\ast, \sigma^2 V^{\ast})$ for each draw of $\sigma^2$. Given the matrix $(I_n - A_M)$ comprises less than $n(m + 1)$ nonzero elements, with each row harboring no more than $m + 1$ nonzeros, the $n \times n$ matrix $(I_n - A_M)^\top D^{-1}(I_n - A_M)$ requires storage of less than $n(m + 1)^2$ elements, and its computational complexity is less than $nm + n(m + 1)^2$. The sparsity observed in the matrix $X^{*\top}X^*$ can be exploited through the use of a Conjugate Gradient (CG) method, setting the stage for an innovative approach in sampling. Specifically, we will solve the linear system $X^{\ast \top}X^\ast \mu^\ast = X^{\ast \top} Y^\ast$ for $\mu^\ast$ using CG, compute $\{a^\ast, b^\ast\}$ and generate posterior samples of $\sigma^2$ from $\mbox{IG}(a^\ast, b^\ast)$. Posterior samples of $\gamma$ are obtained by generating $\eta \sim \mbox{N}(O, I_{n + p}, \sigma^2)$, solving $X^{\ast\top}X^\ast v = X^{\ast\top}\eta$ for $v$ through CG and then obtaining posterior samples of $\gamma$ from $\gamma = \mu^\ast + v$.

This introduction to the application of the Conjugate Gradient method, which we will explore in depth in the subsequent chapter, marks a critical progression in our effort to augment computational efficiency for large-scale spatial models. The capacity of the CG method to operate within these sparse matrix structures not only simplifies calculations but also significantly reduces the computational overhead, making it an indispensable tool in the statistical analysis of extensive spatial datasets. We will detail these advances and their implications for spatial data analysis in the next chapter.

\section{Accelerate Posterior sampling of spatial model through conjugate gradient}

In this section we will test different preconditioning and symbolic decomposition method and compare their performance on accelerating posterior sampling using R.

\subsection{Conjugate gradient}
Consider addressing the challenge of solving a large linear system $Ax = b$, where $A$ is a matrix $n \times n$. Let us presuppose that $A$ is nonsingular with an inverse denoted as $A^{-1}$. For theoretical considerations, assume $A$ to be symmetric and positive definite. This assumption is reasonable because the solution to the system $Ax = b$ aligns with that of $Bx = t$ when $B = A^T A$, $t = A^T b$, with $A^T$ representing the transpose of $A$.

The CG methodology reformulates the solution $Ax = b$ to minimize a quadratic function $\phi(x) = \frac{1}{2} x^T Ax - x^T b$. Starting with an initial guess $x_0$, the CG iterates towards the solution $x^* = A^{-1}b$ within a maximum of $n$ steps. Initially, it sets the search direction $p_0 = r_0 = b - Ax_0$. For subsequent iterations $k = 1, 2, \ldots$, it updates $x_{k-1}$ along $p_{k-1}$ to find $x_k = x_{k-1} + \alpha_{k-1} p_{k-1}$ that minimizes $\phi(x)$ along this direction. The optimal step size $\alpha_{k-1}$ is determined by $\frac{p_{k-1}^T r_{k-1}}{p_{k-1}^T A p_{k-1}}$, and the residual at step $k$ is updated to $r_k = r_{k-1} - \alpha_{k-1} A p_{k-1}$. The next iteration's search direction, $p_k$, is computed using $p_k = r_k + \tau_{k-1} p_{k-1}$, where $\tau_{k-1} = -\frac{r_k^T A p_{k-1}}{p_{k-1}^T A p_{k-1}}$.

The search directions $\{p_0, p_1, \ldots\}$ have been shown to be mutually conjugate and the residuals $\{r_0, r_1, \ldots\}$ are mutually orthogonal, which means $p_i^T A p_j = 0$ and $r_i^T r_j = 0$ for $i \neq j$. The CG method ensures that $x_k$ minimizes $\phi(x)$ within a subspace spanned by $\{r_0, \ldots, r_{k-1}\}$. Given that $r_{k-1} = b - Ax_{k-1}$ represents the negative gradient of $\phi(x)$ at $x_{k-1}$ and indicates the direction of the steepest descent, each CG update is guaranteed to be as effective as or superior to a steepest descent update.

\subsection{Preconditioning method and different solver}
Preconditioning is a transformative concept in computational linear algebra, particularly for iterative methods such as the conjugate gradient method (CG). It involves transforming the original linear system $Ax = b$ into a more tractable form $M^{-1}Ax = M^{-1}b$, where $M$ is a matrix selected to improve the computational efficiency of the system.

The essence of preconditioning lies in the strategic choice of $M$, which is designed to closely approximate $A$ while being simpler to manipulate. This ensures that the preconditioned system maintains the fundamental properties of the original system but is more amenable to iterative solving techniques.

\textbf{Incomplete Cholesky Preconditioner:} This technique is based on an approximate Cholesky factorization of $A$, denoted as $A \approx LL^T$, where $L$ is a lower triangular matrix. The incomplete Cholesky preconditioner, $M_{IC}$, is formed by selectively excluding elements that would result in a denser matrix, thus preserving the sparsity of $A$. When applied to the CG method, the preconditioned system $M_{IC}^{-1}Ax = M_{IC}^{-1}b$ is handled, with $M_{IC}^{-1}$ being more computationally efficient than $A^{-1}$.

\textbf{Identity Preconditioner:} In cases where $A$ is already well conditioned, or when simplicity in computations is paramount, the Identity Preconditioner is employed. Here, $M = I$, the identity matrix, leaving the original system $Ax = b$ unchanged. This approach ensures minimal computational overhead while maintaining the integrity of the original system.

\textbf{Diagonal (Jacobi) Preconditioner:} This method leverages only the diagonal elements of $A$, forming a diagonal matrix $M_D = D$. The resulting preconditioned system $D^{-1}Ax = D^{-1}b$ significantly simplifies the computations as inverting a diagonal matrix is straightforward. It is particularly effective for matrices that exhibit strong diagonal dominance, optimizing computational efficiency without compromising the solution's accuracy.

\textbf{Least Squares Preconditioner:} Aimed at minimizing the residual in the context of the least squares approach, this preconditioner is designed to optimize the solution process. It minimizes the norm of the residual vector, a strategy that proves advantageous in systems where such an approach aligns with the structure and inherent properties of $A$.

Each of these preconditioning techniques offers a unique approach to modifying and solving large linear systems, making them more tractable for iterative methods like CG. By carefully selecting and applying an appropriate preconditioner, computational efficiency can be significantly enhanced, facilitating the resolution of complex linear systems in a more streamlined and effective manner.

Consider the challenge of solving a large linear system $Ax = b$, where $A$ is an $n \times n$ matrix. Assuming $A$ is non-singular, we denote its inverse as $A^{-1}$. For theoretical purposes, we can further presume $A$ to be symmetric and positive definite. This assumption is valid as the solution of the system $Ax = b$ corresponds to that of $Bx = t$ when $B = A^TA$ and $t = A^Tb$, with $A^T$ being the transpose of $A$.

\subsection{Symbolic Cholesky Decomposition}
Symbolic Decomposition emerges as a critical pre-processing step in solving such systems, particularly when $A$ is a large, sparse matrix. This technique focuses on analyzing the sparsity pattern of $A$ without delving into numerical computations. It predicts the structure of non-zero elements in the matrices resulting from decompositions such as LU, Cholesky, or QR, which are pivotal in solving the system.

\textbf{Sparsity Pattern Analysis:} Symbolic Decomposition starts by examining the distribution of non-zero elements in $A$. It systematically identifies the positions of these elements to anticipate the structure of the decomposed matrices.

\textbf{Predicting Decomposition Structure:} Based on $A$'s sparsity pattern, we can predict the non-zero structure in factorized matrices, such as $L$ and $U$ in LU decomposition. This predictive analysis is critical for optimizing computational strategies in subsequent numerical operations.

\textbf{Optimization in Matrix Computations:} The insights gained from Symbolic Decomposition enable efficient memory allocation and computation planning. For example, knowing where non-zero elements are likely to appear in $L$ and $U$ allows for targeted memory usage and minimizes unnecessary computations.

In the context of the CG method, where the focus is on minimizing a quadratic function $\phi(x)$ to solve $Ax = b$, Symbolic Decomposition can significantly enhance efficiency. It aids in optimizing the CG steps by ensuring that the sparsity of $A$ is fully leveraged, reducing the computational and storage burden.

Symbolic Decomposition, in its essence, serves as a foundational step in the efficient handling of large sparse matrices. By meticulously analyzing and predicting the structure of these matrices before actual numerical decomposition, it sets the stage for optimized computational processes, essential in advanced statistical computations involving large linear systems.

\section{Experiments}

\subsection{Simulation settings}
We use a simulation study in this section to discuss the performance of each aforementioned method on different size of matrix. All the algorithm were programmed in R which call the \textbf{Rstan} environment \cite{Stan2025} for building matrix $A_M$ and $D_M$. The conjugate gradient solver, the preconditional method and symbolic decomposition method for sparse linear systems was implemented through \textbf{RcppEigen} \cite{Bates2013}. The nearest-neighbor sets were built using the spConjNNGP function in the spNNGP package. All simulations were conducted on a macOS Sonoma system (version 14.3.1) with 8GM RAM and one Apple M1 processor.

To assess the performance of various conjugate gradient solvers, we conducted tests on large-scale spatial data derived from the dataset described by \cite{Zhang2019}. This dataset comprises 2,827,252 spatially indexed observations of sea surface temperature (SST) collected between June 18 and 26, 2017, spanning oceans from longitudes -140° to 0° and latitudes 0° to 60°. For explanatory variables, we utilized sinusoidally projected coordinates, scaled to units of 1,000 km. Our simulation study evaluated the efficacy of five preconditioning methods, alongside symbolic decomposition and R's default solver, by randomly selecting four subsets of this data, each ranging in size from 1,000 to 1,000,000 observations. We experimented with various conjugate gradient methods under fixed parameters $\phi$ and $\sigma^2$ within a latent nearest neighbor Gaussian process (NNGP) framework, using $m = 10$ nearest neighbors. The settings for $\phi$ and $\delta^2$ were 7 and 0.001, 0.01, 0.1, 1, respectively.

\subsection{Results}
For robust statistical analysis, we replicated each experiment Since solving more replicates on large matrices would take too long, my computer does not have sufficient computational capacity for the task.10,000 , 1,000, 100, and 10 times respectively and compared the total elapsed time. Performance metrics for these methods, applied to matrices of varying sizes, are summarized in the following three tables. The "Relative" column in the table was calculated by dividing the elapsed time of the method under consideration by t
he elapsed time of the fastest method under the same experimental settings.

\begin{table}[hbt!!]
\centering
\begin{tabular}{lll}
\hline
Method & Relative & Elapsed(s) \\
\hline
cgsparse & 1.901 & 7.003 \\
Identity Preconditioner & 2.010 & 7.404 \\
Default solver in R & 1.000 & 3.683 \\
Diagonal Preconditioner & 1.450 & 5.486 \\
Incomplete Cholesky Preconditioner & 7.333 & 27.007 \\
Least Squares Conjugate Gradient & 39.74 & 146.348\\
Symbolic Cholesky Decomposition & 6.377 & 23.487 \\
\hline
\end{tabular}
\caption{Different methods and their performances: On 1000*1000 Matrix for 10000 replications}\label{tab:samplf}
\index{tables}
\end{table}

\begin{table}[hbt!!]
\centering
\begin{tabular}{lll}
\hline
Method & Relative & Elapsed(s) \\
\hline
cgsparse & 1.105 & 7.926 \\
Identity Preconditioner & 1.485 & 10.651 \\
Default solver in R & 2.848 & 20.430 \\
Diagonal Preconditioner & 1.000 & 7.173 \\
Incomplete Cholesky Preconditioner & 4.183 & 30.003 \\
Least Squares Conjugate Gradient & 36.362 & 260.825 \\
Symbolic Cholesky Decomposition & 7.514 & 53.898 \\
\hline
\end{tabular}
\caption{Different methods and their performances: On 10000*10000 Matrix for 1000 replications}\label{tab:sampl}
\index{tables}
\end{table}

\begin{table}[hbt!!]
\centering
\begin{tabular}{lll}
\hline
Method & Relative & Elapsed(s)\\
\hline
cgsparse & 1.073 & 10.829  \\
Identity Preconditioner & 1.391 & 14.037 \\
Default solver in R & & more than 10 mins\\
Diagonal Preconditioner & 1.000 & 10.088 \\
Incomplete Cholesky Preconditioner & 3.592 & 36.234 \\
Least Squares Conjugate Gradient & 4.519 & 45.590 \\
Symbolic Cholesky Decomposition & 14.03 & 141.543\\
\hline
\end{tabular}
\caption{Different methods and their performances: On 100000*100000 Matrix for 100 replications}\label{tab:samp}
\index{tables}
\end{table}

\begin{table}[hbt!!]
\centering
\begin{tabular}{lll}
\hline
Method & Relative & Elapsed(s)\\
\hline
cgsparse & 1.183 &  14.445 \\
Identity Preconditioner & 1.546 & 18.886 \\
Default solver in R &  &more than 5 mins  \\
Diagonal Preconditioner & 1.000 &  12.214\\
Incomplete Cholesky Preconditioner & 5.162 & 63.049 \\
Least Squares Conjugate Gradient & 3.697 &  45.163\\
Symbolic Cholesky Decomposition &  &  more than 5 mins \\
\hline
\end{tabular}
\caption{Different methods and their performances: On 1000000*1000000 Matrix for 10 replications}\label{tab:sam}
\index{tables}
\end{table}

\begin{figure}[hbt!!]
\centering
\includegraphics[width=0.5\textwidth]{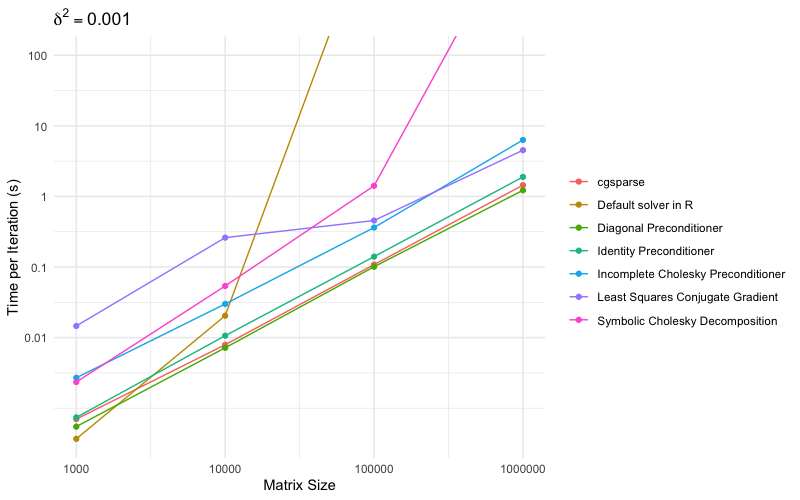}
\caption{Performance of different method:$\delta^2=0.001$}\label{fig:ogo}
\index{figures}
\end{figure}

\begin{figure}[hbt!!]
\centering
\includegraphics[width=0.5\textwidth]{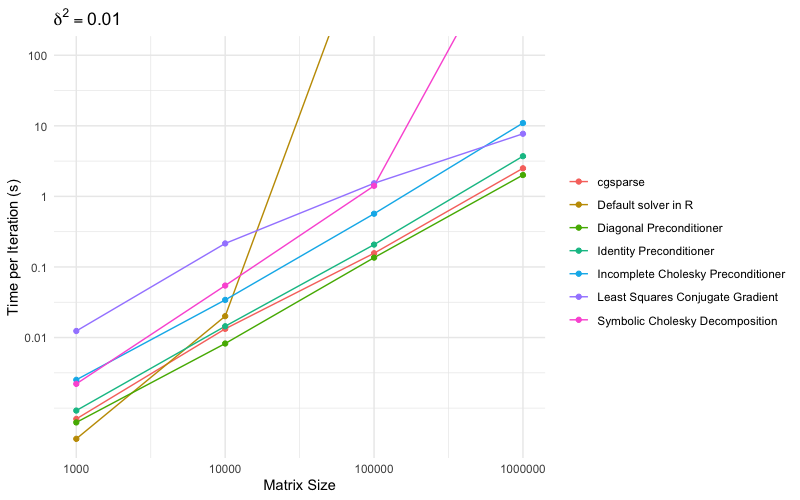}
\caption{Performance of different method:$\delta^2=0.01$}\label{fig:logo}
\index{figures}
\end{figure}

\begin{figure}[hbt!!]
\centering
\includegraphics[width=0.5\textwidth]{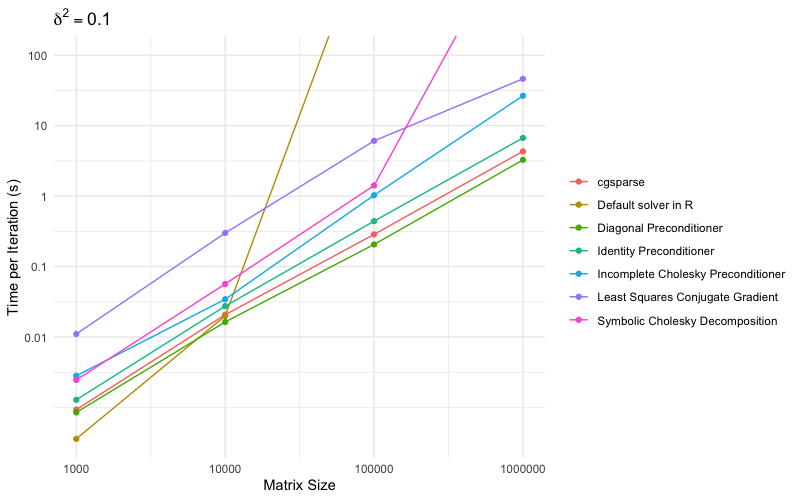}
\caption{Performance of different method:$\delta^2=0.1$}\label{fig:log}
\index{figures}
\end{figure}

\begin{figure}[hbt!!]
\centering
\includegraphics[width=0.5\textwidth]{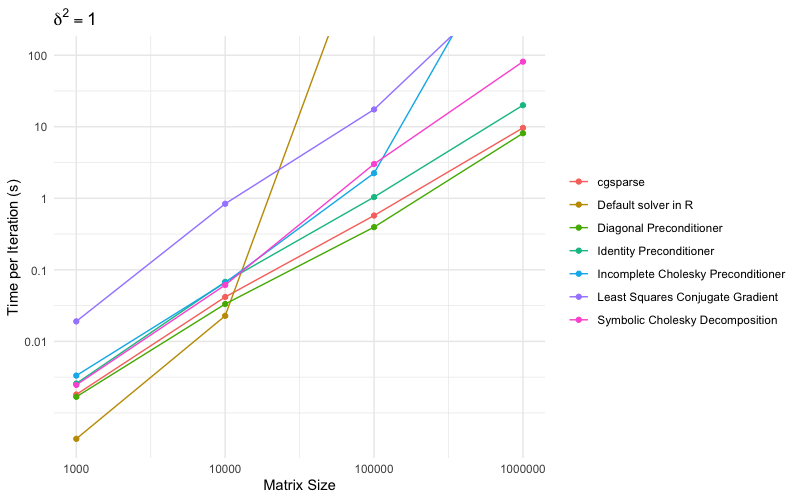}
\caption{Performance of different method:$\delta^2=1$}\label{fig:lo}
\index{figures}
\end{figure}

The Diagonal Preconditioner method outperforms most others by approximately $10\%$ to $20\%$. The default solver in R exhibits superior speed in low-dimensional problems but performs poorly in high-dimensional problems, particularly when the matrix size exceeds 10,000.

Our findings indicate that under low dimensional settings the default solver in R remains significantly faster than its counterparts. However, as we progress to larger matrix sizes, this trend reverses. The scalability challenges of the default solver become apparent, and its efficiency diminishes rapidly when confronted with the computational complexity of larger matrices.

Specifically, the study illustrates that for sample size $1,000$, the default solver outperforms other methods under $\delta^2 = 0.001$. However, its efficiency is not consistent across $\delta^2$; performance degrades as $\delta^2$ increases. When analyzing sample size $10,000$ and above, the solver's limitations are highlighted further, as it fails to complete the computation within a reasonable time frame, particularly for $\delta^2$ values of 0.1 and 1, suggesting an exponential increase in computational burden with increasing matrix size and noise levels.

In contrast, preconditioning methods such as the Diagonal and Incomplete Cholesky show a different pattern. While their performance also deteriorates with increasing matrix size, they maintain a more consistent level of efficiency across different noise levels compared to the default solver. Notably, the Least Squares Conjugate Gradient and Symbolic Cholesky Decomposition methods exhibit a significant reduction in computational speed as both sample size and $\delta^2$ increase, yet their performance decay is more graceful than that of the default solver.

This performance pattern highlights the critical importance of selecting appropriate acceleration methods for different sizes of spatial matrices. Preconditioner and symbolic methods offer viable alternatives to the default solver, especially for large-scale problems where computational resources are a bottleneck. These alternatives not only provide a speed advantage but also enhance the robustness(The absolute difference between corresponding entries in the matrices of the alternative results and the CGSparse results is on the order of $10^{-6}$) of computational processes, ensuring that practitioners can tackle large spatial statistical problems effectively.

The data patterns observed in this study suggest a complex interplay between matrix size, $\delta^2$, and the efficiency of acceleration methods. This emphasizes the need for a dynamic approach in selecting the conjugate gradient solver and its accompanying acceleration technique, one that is informed by both the computational environment and the characteristics of the spatial dataset at hand.

\section{Summary}

This thesis presented a comprehensive simulation study to evaluate the performance of seven acceleration methods for the conjugate gradient algorithm in the resolution of large spatial matrices. The study focused on the use of preconditioner and symbolic decomposition to enhance computational efficiency, crucial for advanced statistical computations in spatial statistics. Our experiments, conducted across a range of sample sizes from $1,000$ to $1,000,000$, were designed to assess the impact of each method under varying levels $\delta^2$.

The empirical results underscored the substantial variance in performance among the acceleration methods, with certain preconditioner, such as the Diagonal Preconditioner, demonstrating consistent computational speed improvements. The symbolic Cholesky Decomposition also showed significant, though variable, efficiency gains, contingent on the matrix size and the measurement noise level. Notably, as the matrix size increased, the relative performance of the preconditioning methods varied, highlighting the necessity for method selection tailored to specific problem dimensions.

In low-dimensional settings, the default solver in R showed remarkable speed, confirming its suitability for smaller matrix sizes. Conversely, for high-dimensional matrices, the solver's performance was suboptimal, reinforcing the value of specialized acceleration methods like the Diagonal preconditioner and cgsparse, which proved to be more robust across varying scales.

The findings from this study not only contribute to a deeper understanding of algorithmic behavior in the context of spatial statistics but also serve as a guiding framework for practitioners in selecting appropriate acceleration techniques for their computational needs. Future research directions could explore the integration of these methods into mainstream spatial statistical software and the investigation of their scalability in distributed computing environments.

The implications of this study are two-fold: Practically, it provides clear guidelines for accelerating conjugate gradient algorithms in spatial statistics, essential for processing large-scale datasets. Theoretically, it lays a foundation for future exploration into the development of new preconditioner and decomposition techniques that can handle even larger datasets more efficiently, paving the way for more profound advancements in the realm of spatial data analysis.

\section*{Acknowledgments}
I extend my deepest gratitude to Professor Lu Zhang for her invaluable guidance and patience throughout the process of writing this paper.

%Bibliography
\bibliographystyle{unsrt}  
\bibliography{main}

\end{document}